\documentclass{svproc}
\usepackage{url}
\usepackage{times}
\usepackage{epsfig}
\usepackage{graphicx}
\usepackage{comment}
\usepackage{algorithm}
\usepackage{algpseudocode}

\usepackage{enumitem} 

\usepackage{footmisc}[bottom]
\usepackage{multicol}
\usepackage{booktabs}
\usepackage{siunitx}
\usepackage{makecell}
\usepackage{multirow}
\usepackage{array}

\usepackage{xspace}

\usepackage{color}

\makeatletter
\DeclareRobustCommand\onedot{\futurelet\@let@token\@onedot}
\def\@onedot{\ifx\@let@token.\else.\null\fi\xspace}

\def\etal{\emph{et al}\onedot}
\makeatother

\begin{document}
\mainmatter              
\title{Smart mobile microscopy: \\ towards fully-automated digitization}
\titlerunning{Smart mobile microscopy}  
%
\author{Anastasiia Kornilova\inst{1} \and Iakov Kirilenko\inst{2} \\
Dmitrii Iarosh \and Vladimir Kutuev \and Maxim Strutovsky}
\authorrunning{Anastasiia Kornilova \etal} 
%
\tocauthor{Anastasiia Kornilova, Iakov Kirilenko, Dmitrii Iarosh, Vladimir Kutuev, Maxim Strutovsky}
\institute{Skolkovo Institute of Science and Technology, Russia\\
\email{anastasiia.kornilova@skoltech.ru},
\and
Saint Petersburg State University, Russia\\
\email{y.kirilenko@spbu.ru}
}

\maketitle              

\begin{abstract}
Mobile microscopy is a newly formed field that emerged from a combination of optical microscopy capabilities and spread, functionality, and ever-increasing computing resources of mobile devices. Despite the idea of creating a system that would successfully merge a microscope, numerous computer vision methods, and a mobile device is regularly examined, the resulting implementations still require the presence of a qualified operator to control specimen digitization. In this paper, we address the task of surpassing this constraint and present a ``smart'' mobile microscope concept aimed at automatic digitization of the most valuable visual information about the specimen. We perform this through combining automated microscope setup control and classic techniques such as auto-focusing, in-focus filtering, and focus-stacking~--- adapted and optimized as parts of a mobile cross-platform library.

\end{abstract}

\section{Introduction}

With the development of optical microscopy technologies, simple microscopes' cost has become low enough for their mass usage. This breakthrough opened up opportunities for the development of mobile microscopy~--- the field where smartphone camera and computational resources are used in combination with universal optical microscopes for affordable data collection and diagnostics in different areas: disease diagnosis in telemedicine~\cite{ScreeningTool}, agriculture analysis~\cite{Bogoch2013}, water and air quality~\cite{Wu2017}, bacteria and virus detection far from professional equipment~\cite{FOZOUNI2021323}~\cite{Koydemir2015}.

Despite a significant number of proposed solutions and designs for mobile microscopes over the past ten years~\cite{Kuhnemund2017}~\cite{ClinicalMobile2009}, the field of mobile microscopy still requires the presence of an expert to digitize specimens in focused zones for further study. This generates demand for a ``smart'' mobile microscope that will be able to automatically extract essential and comprehensive information about specimen. The possible solution could be to use automated scanning over Z-axis and along XY-plane to move the stage and provide a control system with feedback loop, which implements automatic smartphone image processing, to extract focused planes with full coverage of the specimen. The closest solution to this idea is proprietary CellScope device~\cite{CellScope}, which, as mentioned in~\cite{CellScopePatent}, provides scanning over Z-axis and XY-plane, automated with motors, and software for iOS devices (iPhones or iPads) that supports controlling microscope with manual gestures. Authors also tackle the problem of software control ``intelligency'' and highlight possible features that can enhance user experience: autofocus, 3D-reconstruction, automated detection of bacterias, but do not provide enough discussion of used methods and implementation details.

In our research we consider the next main features of ``smart'' mobile microscope for Z-axis scanning that could help to avoid presence of an expert in digitizing the most essential information about the specimen.

{\small\begin{figure}
\begin{center}
  \includegraphics[width=0.8\linewidth]{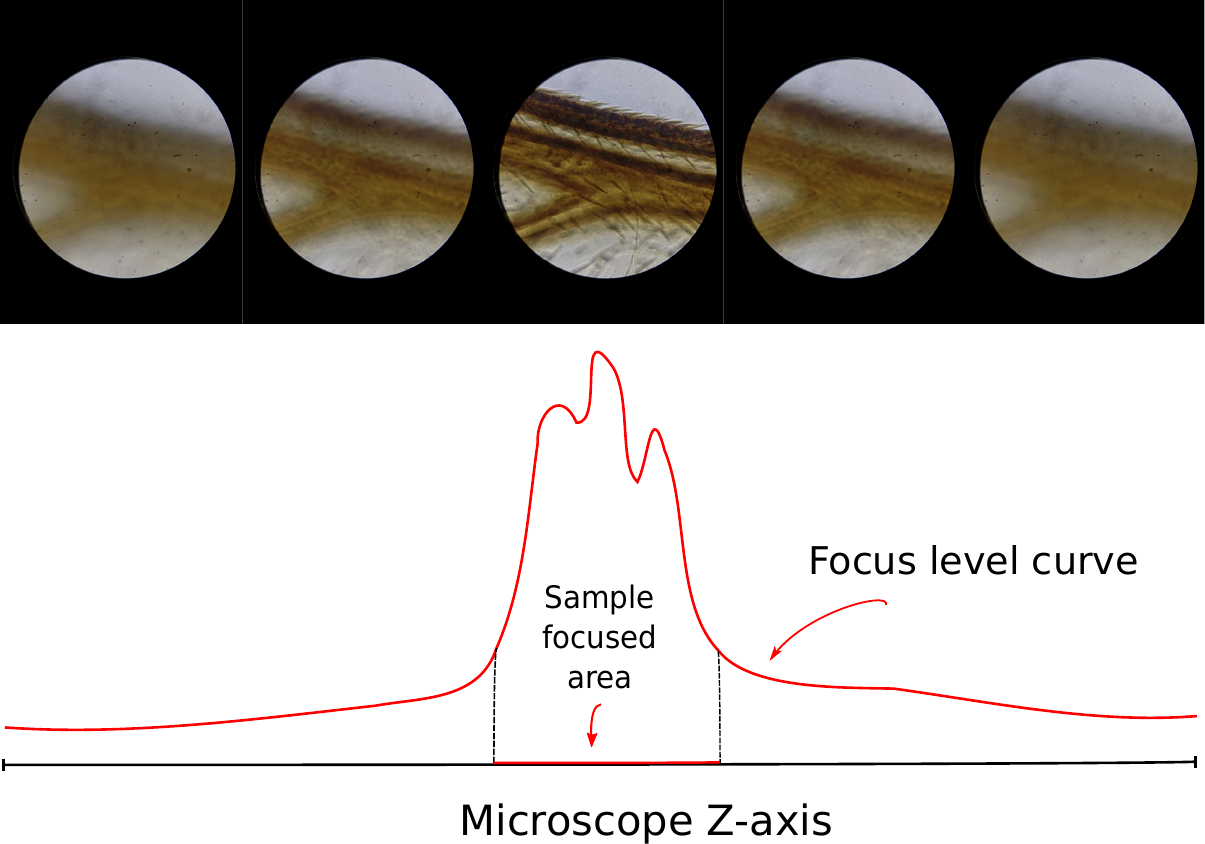}
\end{center}
\caption{Demonstration of Z-stack images with respect to focus measure. The most valuable information about specimen is contained in sample focused area which is usually less than 0.1 part of the stack length.}
\label{basic_focus}
\end{figure}
}

{\small\begin{figure}
\begin{center}
  \includegraphics[width=0.6\linewidth]{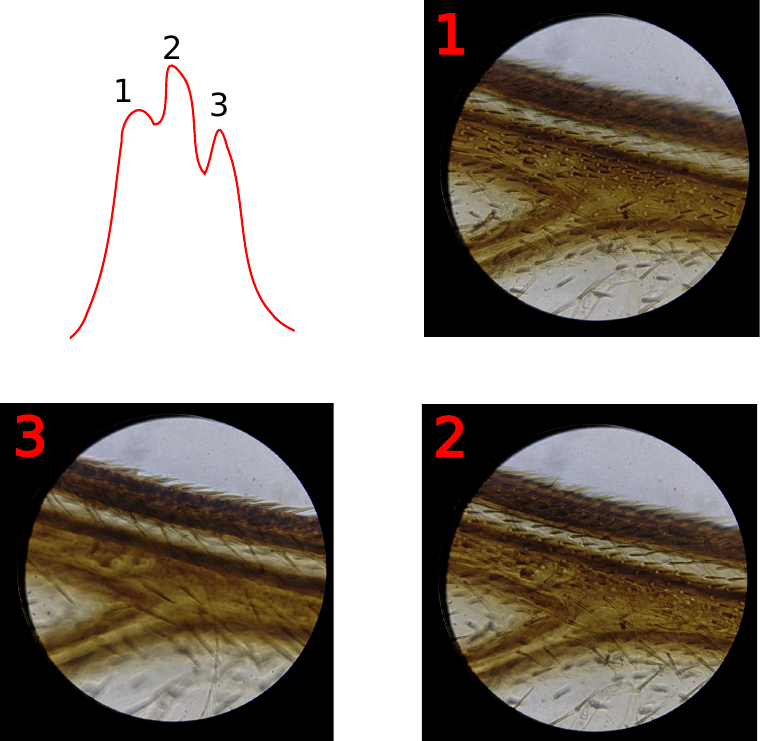}
\end{center}
\caption{Example of sample, the full focus coverage of which is a subset of 3 images from Z-stack with different focused zones. All those images correspond to the local peaks in the sample focus area.}
\label{several_focuses}
\end{figure}
}

\textbf{Fast search of sample focused area.} Usually, a segment of Z-axis, where all the essential information about the sample is concentrated, is more than 10 times less than the length of Z-axis (schematically depicted on Figure~\ref{basic_focus}). Absence of the method that can briefly detect that segment requires to scan the whole Z-axis slowly in order not to get blurred images because of rolling-shutter effect of CMOS smartphone cameras. This process, crucial to the study of the specimen, could be drastically optimized by a fast passage along Z-axis followed by an approximate detection of the Z-axis area containing sample focused parts and an analysis of only that small segment.

\textbf{Full focus coverage and in-focus filtering.} The most essential information about specimen could be described as the full focus coverage~--- a subset of unique Z-stack images that contains all focused zones of the specimen as depicted in Figure~\ref{several_focuses}. To make it time-efficient, the smart mobile microscope should first define a borders of sample focused area.

\textbf{Focus stacking (optional).} Also, we optionally consider a sub-task of focus stacking the purpose of which is to combine all images with different focus zones to get an all-in-focus image that could be useful for an operator to verify preliminary results of scanning. 

Notwithstanding a huge variety of methods both in classical computer vision and in actively developing field of Deep Learning, mobile microscopy case deserves special attention because of the next two reasons: (i)~mobile microscopy specifics~--- bright-field mode, lower quality of images in comparison with professional microscopes, artifacts in optical system (dust, water drops, condensate) and (ii)~lower smartphone performance in comparison to PCs, which raises additional requirements to algorithms in terms of near real-time response and processing.

In this paper we discuss existing implementations of the above-mentioned features for professional microscopes, their computation limitations for usage in a ``microscope-smartphone'' system, as well as provide quality estimation of methods and performance estimation on target devices (smartphones), propose proper modifications of the method to achieve trade-off between performance and quality that is crucial for near real-time processing. Also, we present a design for a cross-platform library that uses the discussed methods and, without affecting the generality of cross-platform application, demonstrates its effective embedding on the devices with Android OS.

The paper is organized as follows: in section 2 we discuss existing work in the field of mobile microscopy, both in the field of devices and algorithms, as well as existing work regarding the highlighted sub-tasks in classical microscopy. Section 3 contains information about the mobile microscope, that we use in this paper, and describes mobile microscopy imaging specifics. Section 4 covers the main methods we propose for fast peak search, extraction of full focus coverage and focus stacking. In section 5 we discuss the design and architecture of our cross-platform mobile microscope library which was developed to be performance-effective and available on all the variety of modern smartphones. Section 6 contains qualitative results of proposed methods and their time performance in order to prove their near real-time performance on smartphones.
\section{Related works}
\subsection{Mobile microscopy}

In the past decade mobile microscopy has demonstrated its effectiveness in different fields for fast diagnostics.

In 2009, authors of~\cite{ClinicalMobile2009} demonstrated a potential application of mobile microscope for clinical use in detection of P. falciparum-infected and sickle red blood cells and M.~tuberculosis-infected sputum samples. In 2013, the work~\cite{Bogoch2013} presented a proof-of-concept study for a mobile microscope use in soil-transmitted helminth infection diagnostics: Ascaris lumbricoides, Trichuris trichiura and hookworm. Authors of \cite{Koydemir2015} demonstrated successful sample imaging on a mobile microscope for automatic detection of Giardia lamblia cysts and proposed an algorithm that could be run on a server side. In 2017, three impactful researches were released:~\cite{ScreeningTool} successfully proposed to apply an automated tablet-based mobile microscope for oral-cancer screening with brush biopsy,~\cite{Wu2017} presented design for a mobile microscope platform c-Air and a machine learning algorithm for monitoring air quality indoors and outdoors,~\cite{Kuhnemund2017} demonstrated that targeted next-generation DNA sequencing reactions and in situ point mutation detection assays in preserved tumor samples can be processed with a mobile microscope. 

Moreover, mobile microscopy can be used for virus detection. In 2013, authors of~\cite{Wei2013} presented a compact solution for imaging of viruses with mobile phone. They proved that even such small particles can be observed in the field conditions using mobile microscopy. Such mobile microscope systems become very useful in the time of wide-spread virus diseases when medical centers tend to be overwhelmed or situated too far from the source of a disease. Application of this ability was demonstrated by authors of~\cite{Chung2021}. They presented a solution for detection of Norovirus using a custom-built smartphone-based fluorescence microscope and a paper microfluidic chip. Another example of virus detection mobile microscopy success was presented by authors of~\cite{FOZOUNI2021323}. They created a SARS-CoV-2 detection method using mobile microscopy. It enables rapid, low-cost, point-of-care screening for SARS-CoV-2. The test is also fast, accurately identifying SARS-CoV-2-positive clinical samples with a measurement time as low as 5 minutes.

Besides clinical usage and algorithms, a lot of different designs and constructions were suggested to support different modalities, to improve quality of a mobile microscope optical system both in direction of special lenses~\cite{Pechprasarn2018}~\cite{Kim2015} and external devices~\cite{Jung2017}~\cite{Meng2016}.

\subsection{Auto-focus and image filtering}

Auto-focusing is a well-known topic which is essentially researched during the development of the most optical systems such as digital cameras, DSLRs, bright-field microscopes, etc. The essential step in both auto-focus and filtering methods is extracting focus-related information, which is commonly achieved by applying focal measure operators. The multitude of the operators can be classified as follows: based on the gradient, based on the Laplace operator, based on wavelet transforms, statistical, mixed. A detailed overview of such operators was provided by Pertuz~\etal~\cite{MeasureOperators}. Also, the authors compare the performance of operators in various aspects (noisy images, different size of the operator window).

The use of focal measure operators for auto-focus in microscopy has been studied by O. A. Osibote~\etal~\cite{BestOperators}. The authors propose to use focal measure operators for all frames taken by the microscope and to take the frame with maximum value as the most focused position. Also, the authors compared the operators of the focal measure in terms of the quality of work, accuracy and speed of finding focus with their help in the field of microscopy. Based on the study, it was concluded that the most suitable operators for application in this area are: the normalized variance, the Brenner gradient, the sum-modified Laplacian, the energy of the Laplacian, Vollath F4 and the Tenengrad operator.

The task of auto-focusing is attempted to be solved by alternative means as well. Bueno-Ibarra~\etal~\cite{Ibarra2005} presented an algorithm based on a one-dimensional Fourier transform and the Pearson correlation and estimated its applicability to the field of automated microscopes. Another approach is application of deep convolutional neural networks, e.g. for detection of focused frames as described by Yang~\etal~\cite{Yang2018}. Dastidar~\cite{Dastidar2019} proposed an auto-focusing method based on CNNs. The presented method shows superior performance and is claimed to be ideal for deployment in a device with limited computing resources, yet the presented hardware requirements are still hardly matched by mobile devices and the measured processing time is too high for an interactive workflow. 

Auto-focusing methods also prove themselves useful in image filtering. Vaquero~\etal~\cite{Vaquero2011} presented an algorithm capable of selecting a minimal set of images, focused at different depths, such that all objects are in focus in at least one image. The method was enhanced by Choi~\etal~\cite{Choi2017} and adjusted for use in non-mobile devices. An alternative approach is presented by Li~\etal~\cite{Li2018} as a part of a round-trip scene-adaptive image acquisition system.
 
The problem of the dirt presence in images is that it is usually found on the instrument glass or microscope lens, therefore these areas might be in focus while the main specimen is blurred. As a result, a frame may be falsely identified as in focus due to the presence of such dirt in focus in the frame. One possible solution to this problem is suggested by V. Hilsenstein~\cite{Dust}. The author takes into account the peculiarities of dirt relative to the sample (its shape and size) and provides filters capable of detecting objects that we classify as dirt in photographs and removing such frames from the original set.

\subsection{Focus-stacking}

Focus-stacking, or multi-focus image fusion, is well-developed in the field of standard image processing, being presenting as a huge variety of methods both in classical computer vision and in Deep Learning area. With respect to Sigdel~\etal~\cite{Harris} classification, algorithms based on classical computer vision could be divided into 3 groups: pixel-based approaches~\cite{Liu2001}, neighbor-based approaches~\cite{Xiangzhi2015,Li2013,Liu2015}, and wavelet-based approaches (usually based on wavelet transformations)~\cite{Tian2012}. Deep learning approaches are significantly superior to the standard ones, when compared qualitatively, both with CNN architectures~\cite{Zhang2020,Liu2017,Tang2017} and with GAN architectures~\cite{Genga2020}. Nevertheless, mobile hardware generally lacks the processing power for efficient utilization of the neural network. That is why in the mobile microscopy pipeline, we tend to use classical approaches.

\section{Background}
\textbf{Mobile microscope description.}
In our work we use a mobile microscope provided by the sponsor company. The microscope scheme is depicted on the Figure~\ref{microscope_scheme}. The microscope consists of a base on which a linear rail guide is attached to move stage with the specimen along Z-axis. The movement along the guide is provided by an actuator driven by a stepper motor JK42HS28-0604-02AF. To control motor and communicate with smartphone via Bluetooth, Arduino~Uno~R3 board is used. This configuration provides opportunities to set movement speeds that could be exploited to speed up the process of specimen study. To connect a smartphone to the eyepiece independently of its model, special holders are used.

{\small\begin{figure}
\begin{center}
  \includegraphics[width=0.5\linewidth]{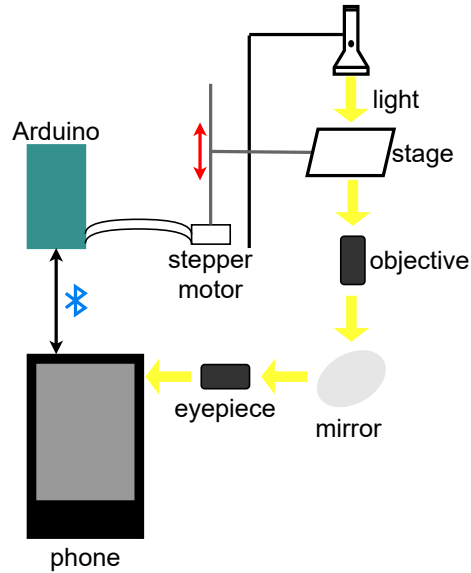}
\end{center}
\label{microscope_scheme}
\caption{Microscope scheme.}
\end{figure}
}

\textbf{Mobile microscopy specifics.}
Before attempting to solve discussed issues for the smart mobile microscope, it is important to state features specific to the field of mobile microscopy that could heavily influence the solution design. The solution should use only lightweight methods as the interactive usage of a mobile platform implies low response time. The focal stack is captured by moving a stage, therefore one should consider the time required to set a specific z-position. As a consequence, random frame retrieval is almost as time-consuming as a full range scan. Since a mobile microscope is rarely used in perfect conditions, there is a high possibility of facing numerous defects such as: dust, optical effects, jolting and low image resolution. A mobile microscope is a multi-purpose device, hence it is unlikely that any assumptions regarding properties of the specimen would hold, except the most general ones.

\section{Methods}

\subsection{Fast search of sample focused area}

A number of possible z-positions greatly exceeds the thickness of most of the samples in a mobile microscopy case. Therefore a naive image stack gathering approach via moving an optical lens with a fixed step-size yields a formidable number of frames, most of which do not capture any area of specimen surface in focus. As focus stacking and advanced image selection methods performance suffers significantly on large-sized frame stacks, the following task has emerged. Let $\{s_1, \dots, s_n\}$ be the aforementioned Z-stack and $\{s_{k_1}, \dots, s_{k_m}\}$ be the frames required for full focus coverage. Then find indices $i, j$ such that $[i,j] \supset \{k_1, \dots, k_m\}$ and $|j - i| \rightarrow min$. Informally, the goal is to detect a minimal segment of the optical lens positions that will contain a full focus coverage set.

\begin{figure}[t]
\centering
\includegraphics[width=\linewidth]{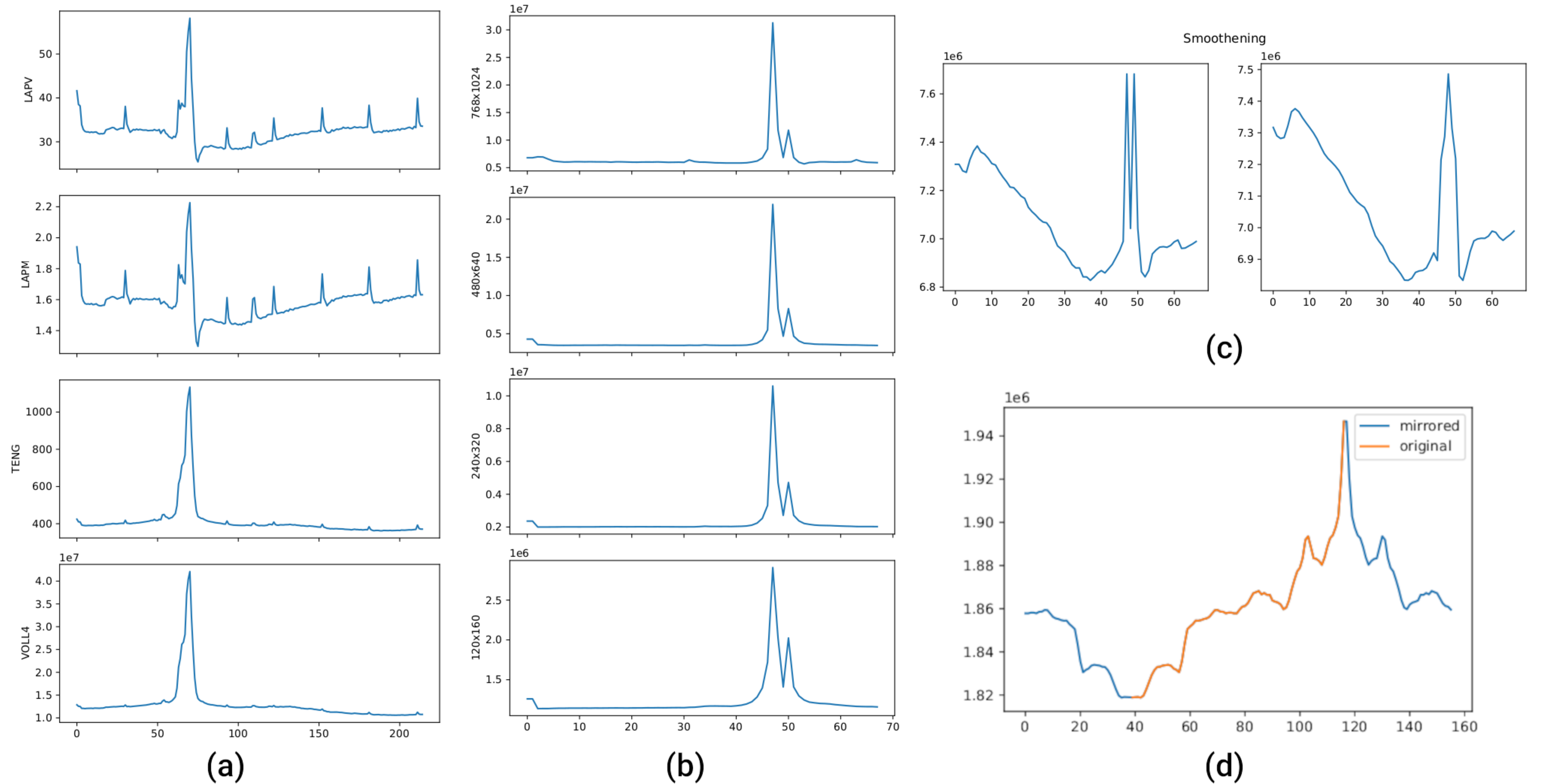}
\caption{
\textbf{(a)} Focal curves obtained by using each of the four evaluated FM operators on a single z-stack. X-axis depicts frame number, y-axis~--- FM value. Considering the graph, it is tempting to choose the VOLL4 operator since the curve contains the least amount of false peaks.\\
\textbf{(b)} Focal curves obtained by altering the z-stack image resolution. One can see that reducing the resolution does not flatten the distinguishable peak. \\
\textbf{(c)} An exemplary case illustrating the necessity of the smoothening step. Here, avoiding this step could lead to a recognition of only half of the peak.\\
\textbf{(d)} An illustration of the mirroring step. As the prominence of near-end local maxima (such as the one at $\sim\,$120 frame) evaluates to near-zero values, despite the peak being empirically the most prominent one, we have decided to artificially extend the curve, eliminating such an inconsistency.  
}
\label{peak_search_graphs}
\end{figure}

The underlying idea is to capture a low-resolution sparse Z-stack (to achieve higher performance in processing) to retrieve a low-quality focal curve and detect the most prominent peak. The boundary indices of the detected peak frames would constitute the demanded segment. The implementation of the concept was held in the next three steps.

\textbf{Step 1.} Select the most appropriate focus measure (FM) operator, i.e., a lightweight operator which yields a smooth focal curve. As it was stated, the algorithm needs to cause the system to stall because of high-demanding computations as little as possible. Consequently, the operator implementation should complete the evaluation of a single frame focus before the next frame is captured. Therefore, two fundamental parameters were considered: (i) performance of implementation and (ii) focal curve parameters critical for peak detection. In our work, we consider the next focus measure operators LAPM, LAPV, TENG, and VOLL4~\cite{Vollath1987} whose performance in microscopy was thoroughly evaluated and proven previously~\cite{BestOperators}.

FMs performance measurements on the target device are presented in Table \ref{time_cmp_obj_func}. For every specimen from the dataset, a graph with four focal curves corresponding to each of the FMs was plotted for a side-by-side empirical comparison as presented in Figure~\ref{peak_search_graphs}.a.

\begin{table}
\centering
    \caption{Performance of focus measure operators with respect to image resolution, ms. CI=0.95}
    \begin{tabular}[c]{
    |S[table-format=4.4,output-decimal-marker=\times]
    *4{|S[table-figures-uncertainty=2, separate-uncertainty=true, table-align-uncertainty=true,
          table-figures-integer=3, table-figures-decimal=2, round-precision=2,
          table-number-alignment=center]}
    |}
    \toprule
        \multicolumn{1}{|c|}{Resolution} & \multicolumn{1}{c|}{$TENG$} & \multicolumn{1}{c|}{$LAPM$} & \multicolumn{1}{c|}{$LAPV$} & \multicolumn{1}{c|}{$VOLL4$} \\\hline
        1920.1080 & 406.23 \pm 0.94 & 134.06 \pm 0.35 & 207.45 \pm 0.42 & 115.86 \pm 0.34 \\ \hline
        1024.768  & 145.0 \pm 0.47  & 39.68 \pm 0.1   &  52.79 \pm 0.24 & 28.33 \pm 0.1 \\ \hline
        464.848   & 70.57 \pm 0.2   & 19.86 \pm 0.01     & 32.75 \pm 0.1   & 12.7 \pm 0.04 \\ \hline
        640.480   & 51.10 \pm 0.2   & 14.70 \pm 0.1 & 24 \pm 0.10   & 10.95 \pm 0.04 \\ \hline
        160.120   & 2.4 \pm 0.02    & 0.67 \pm 0.01      & 0.92 \pm 0.01      & 0.48 \pm 0.01 \\
        \bottomrule
    \end{tabular}%
    \label{time_cmp_obj_func}
\end{table}

Considering the experimental results, the VOLL4 FM was selected, being the fastest and yielding focal curves suitable for peak detection.

\textbf{Step 2.} Determine the lowest resolution and the highest fixed-size step that would still be sufficient to find the peak. We are interested in these extreme parameters as lower resolutions and higher step-sizes would lead to the decreased computational complexity of processing the stack. To achieve this goal, we have used the VOLL4 FM to plot focal curves for (i) 4 distinct step-sizes easily available in our setup and (ii) 5 different resolutions standard for wide variety of smartphones.

The majority of available specimens were used in this experiment. A side-by-side empirical examination of the gathered graphs (as shown in Figure~\ref{peak_search_graphs}.b) resulted in a conclusion that the highest step-size and the lowest resolution are adequate for the peak detection. 

\textbf{Step 3.} Apply the existing peak detection techniques to find the focus peak on the generated focal curve. In our work, we have used the following two-steps algorithm: (i) find all local maxima by simply iterating over the values array and then (ii) pick out only those peaks that meet additional criteria, for example, the most prominent peak. Therefore this step has linear complexity, and could be easily implemented using standard signal processing techniques and convolutions.
\makeatletter
\algrenewcommand\ALG@beginalgorithmic{\small}
\makeatother
\begin{algorithm}[t]
  \caption{Fast search of sample focused area}
  \begin{algorithmic}[1]
      \State $image\_stack \gets fast\_scan$
      \State $focal\_curve \gets$ \Call{voll4}{$image\_stack$}
      \State \Call{smoothen}{$focal\_curve$}
      \State \Call{mirror}{$focal\_curve$}
      \State $x, w \gets$ \Call{bin\_search\_prominent\_peak}{$focal\_curve$}
      \State \Call{map\_back}{$x, w$}
      \State \Return $x, w$
  \end{algorithmic}
\label{peak_search_alg}
\end{algorithm}

Having discussed the prerequisites, we are ready to present the algorithm~\ref{peak_search_alg} workflow. First, we obtain a coarse Z-stack, i.e., by selecting low resolution and high step-size, of a specimen. Then, we use a VOLL4 FM to acquire a focal curve of the stack. After that, we smooth a focal curve to reduce z-stack defects (shown at Figure~\ref{peak_search_graphs}.c) and extend the left and the right ends of the curve by mirroring its first and second half, respectively. The second step of the described preprocessing is essential for the correct near-end peak detection (see Figure~\ref{peak_search_graphs}.d). As the curve is ready for maxima analysis, we perform a binary search for such a \texttt{prominence} value that would result in a single peak. Finally, we map the peak coordinate back to original array indices and combine it with its width to return the requested segment.

\subsection{Full focus coverage}
After the necessary segment is detected by the fast search of the focus sample zone algorithm, the smart mobile microscope software should extract the full focus coverage set containing high-quality, unique images with different zones in focus. In mobile microscopy, this task is complicated with artifacts caused by imperfect optical systems and recording conditions: dust, water drops or dirt, jolting, illumination alteration. To get perfect slices with different focused zones without any distortions, it is necessary to scan slowly inside the focus sample zone, which requires filtering of near-duplicate images. Assuming these specifics, the next requirements were formulated for a full focus coverage extraction algorithm: (i) minimize the image stack size while keeping for every sample surface area at least one frame that contains its focused image; (ii) remove completely blurred frames; (iii) remove frames including images of dust or dirt.

\begin{figure} [t!]
\centering
\includegraphics[width=\linewidth,clip,trim=0cm 5cm 0cm 5cm]{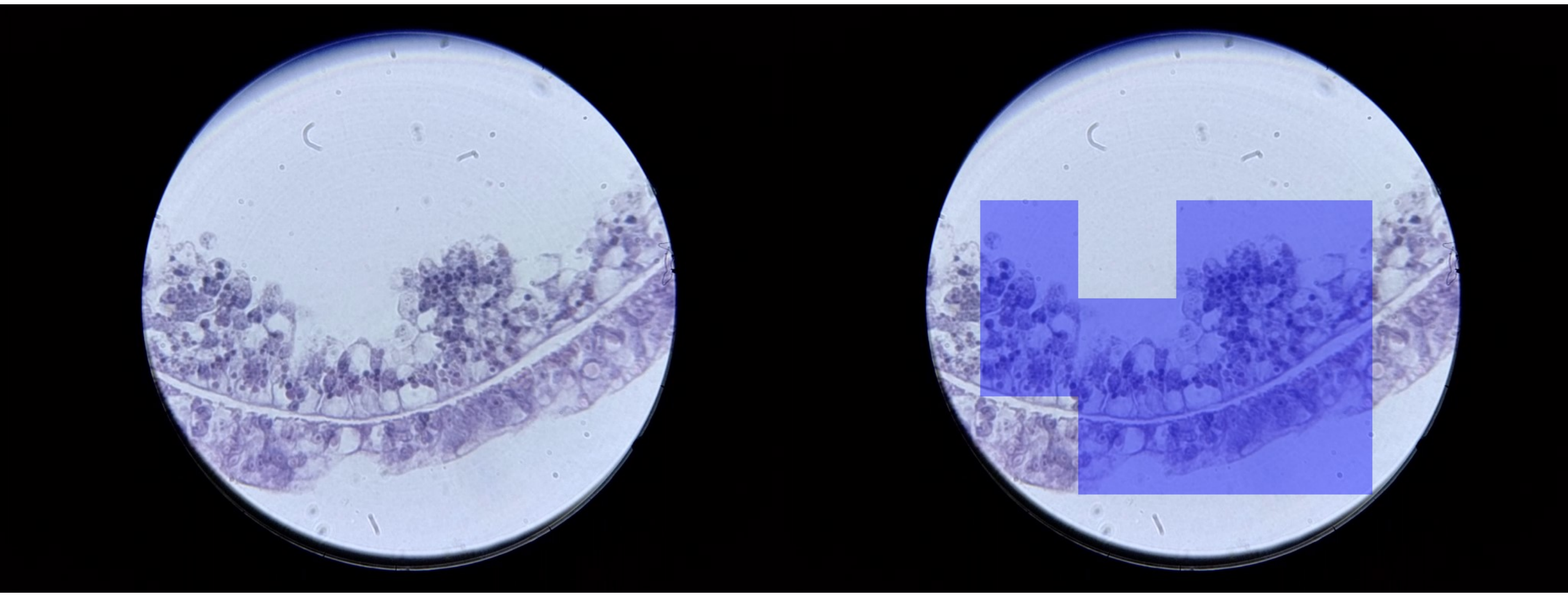}
\caption{An example of image division to sectors for focus measuring.}
\label{Sectors}
\end{figure}

As it was discussed previously, the most reliable and time-tested way for extraction focused frames is to use a focus measure operator, which in comparison to DL approaches is more performance-effective, which is crucial for near real-time processing. This approach yields relevant results in selecting the most focused frame. Still, it does not cover correctly other Z-stack images containing important information in focus in a small part of the image as it was demonstrated in Figure~\ref{basic_focus}. To solve this issue, we follow idea from image processing field~\cite{Vaquero2011,Choi2017} and use a sector-based focus measure operator, which means that the image is divided into sectors by grid and for each sector focus measure is calculated. Such division allows not to lose even small focused details. An example of division into sectors is demonstrated on Figure~\ref{Sectors}.

Because smartphone optical system has a small focal length, images obtained from the mobile microscope are usually framed with dark corners. Such an effect could affect correct frame selection in the sector-based approach because some sectors may contain nearly all pixels in those zones. To avoid this issue, we use the mask obtained from the binary threshold and do not consider images with more than half of the pixels marked by this mask. To solve the issue with near-duplicate frames because of the slow scanning, we use a threshold for the pixel-based difference to drop similar frames.

\begin{figure} [t!]
\centering
\includegraphics[width=0.6\linewidth,clip,trim=5cm 4.3cm 8cm 5cm]{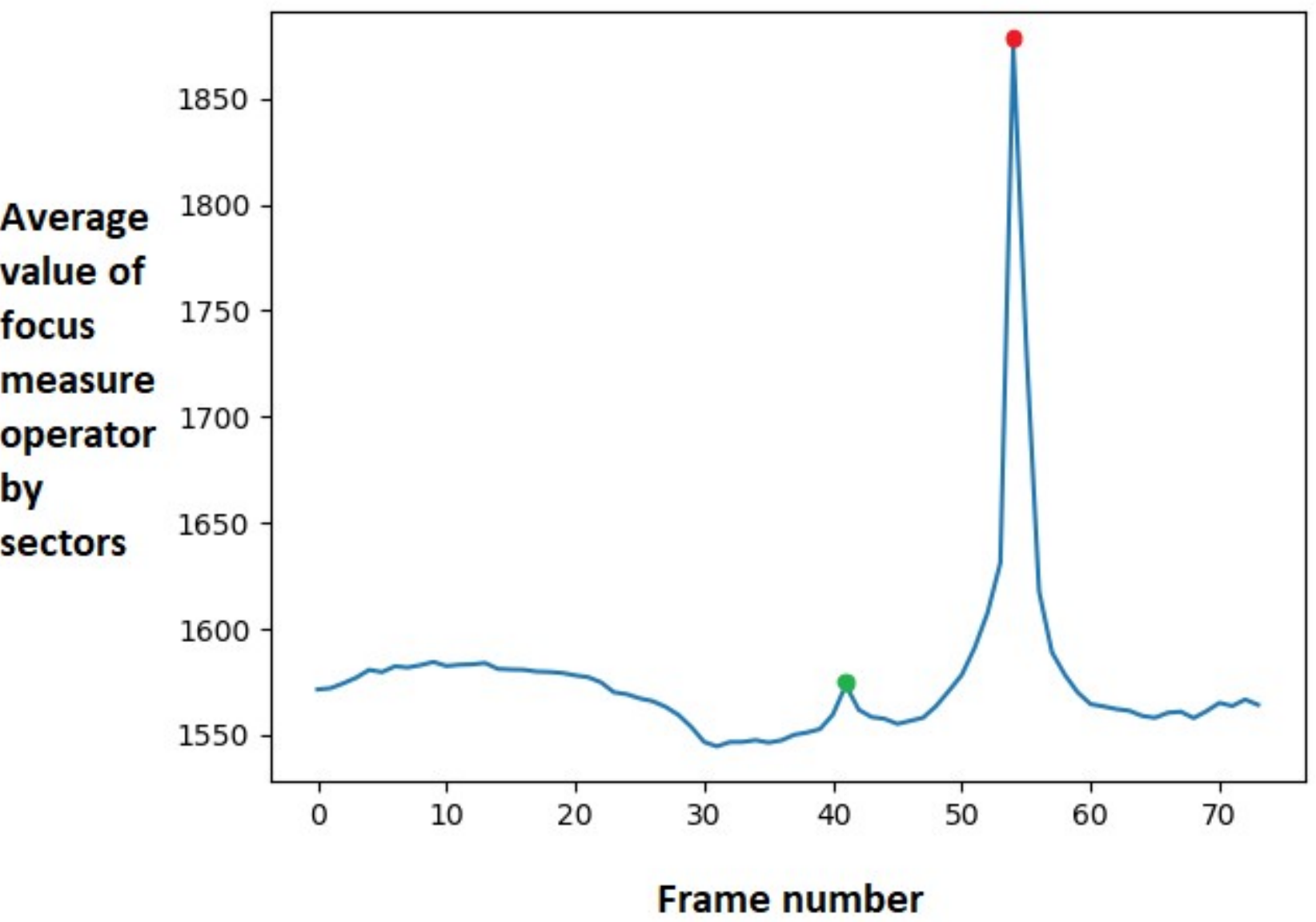}
\caption{Focus measure for every frame in Z-stack. Red point demonstrates the most prominent peak, the green one~--- related to dirt image.}
\label{GraphFM}
\end{figure}

As discussed, it is also essential to filter out false ``in-focus'' frames containing mobile microscopy artifacts: dirt and condensate. To detect those frames, we follow the following observations of the focus measure operator for the whole Z-stack frames: dirt usually fills only a small part of the frame, and when it is in focus, the sample which is situated far from the dirt on the vertical axis is always blurred. According to this, frames with the sample in focus are detected as massive peaks on the focus measure curve, whereas dirt peaks are small ones far enough from the prominent peak as in Figure~\ref{GraphFM}.

\subsection{Focus-stacking}
Focus-stacking is a technique of generating a single all-in-focus image from a full focus coverage set that could help the operator verify the scanning results visually. If every part of the focus-stacked image is well-defined, then the specimen digitizing was done correctly.

We consider existing classes of algorithms (pixel-based, neighbor-based, wavelets, and DL models) for focus stacking from the performance point of view for implementing this functionality for mobile microscope, keeping in mind near real-time requirements for processing on the smartphone. The basic comparison of algorithms in terms of performance, quality, and stability (how it performs when the input stack's size is increased) are presented in Table~\ref{FSTime}. In our work, in mobile microscope library we implemented the next representative of every class besides DL approach: pixel-based with TENG operator~\cite{BestOperators}, neighbor-based~\cite{Harris}, wavelets~\cite{ComplexWavelet}.
\begin{table}
\centering
\caption{Focus-stacking algorithms characteristics for approaches}
\begin{tabular}[c]{|l|c|c|c|c|}
\hline
 & \thead{Performance} & \thead{Quality} & \thead{Stability} \\
\hline
Pixel-based & high & low & low\\
\hline
Neighbor-based & medium & medium & medium\\
\hline
Wavelets & low & high & high\\
\hline
DL models & super-low & high & high\\
\hline
\end{tabular}
\label{FSTime}
\end{table}

\section{Software}
\textbf{Cross-platform library.}
To make the library available on the most popular smartphone OS Android and iOS, C++ language was used, which in comparison to cross-platform frameworks Xamarin~\cite{xamarin}, React Native~\cite{react}, Flutter~\cite{flutter} allows to compile the library for the target architecture and use it both in these frameworks and in native Android and iOS applications. Besides the majority of considered algorithms uses common computer vision techniques~--- convolutions and different filters~--- therefore the OpenCV~\cite{OpenCV} library was used. It provides not only cross-platformity among considered OS but also ARM NEON for the main library methods to be performance-effective on smartphones. 

To use the library in an Android app, Java Native Interface (JNI) could be used. In case of iOS platform, usage of the library from the app written on Swift is carried out by Objective-C layer that calls C++-methods of the library. To avoid support of both platforms with additional layers between app and library, Djinni~\cite{Djinni} library is used to generate cross-language type declarations and interface bindings to connect C++ with either Java or Objective-C. 
The wide variety of smartphone architectures requires the use of different compilers. To automate this process our build system uses Polly tool~\cite{Polly}~--- collection of scripts and build-files for different build tools.

\begin{figure*}
\centering
\includegraphics[width=\textwidth]{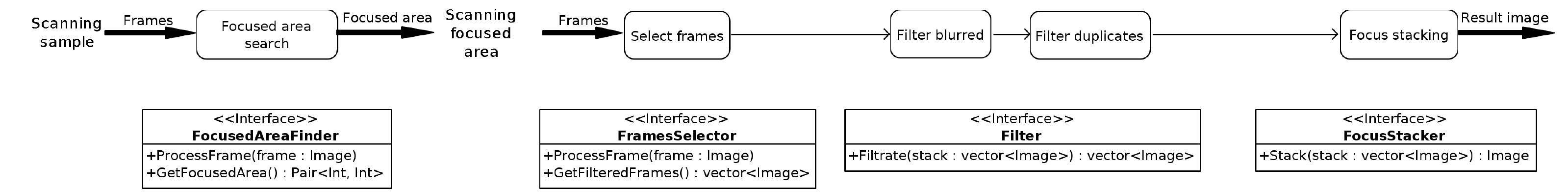}
\caption{Library interface architecture with respect to processing pipeline}
\label{pipeline}
\end{figure*}

\textbf{Processing pipeline and architecture.}
Real-time processing, or post-processing of the Z-axis stream, could be formulated as a pipeline containing different modules to perform the frames' processing. To natively follow this process, the library implements Pipes\&Filters architecture. The overview of the library modules is presented in Fig.~\ref{pipeline}. Every action is represented by a particular module that takes a set of frames as an input and then necessarily processes them and returns the result set to the next module in the chain. Every module is an interface and a set of implementations that can be changed during the pipeline configuration. Also, such an architecture can be easily extended for the usage of new modules or reconfigured for disabling some unnecessary for current launch modules. Except for simple automatic scanning of the sample, other operations can be done using this library as a base.

\section{Experiments}
For experiments, dataset from mobile microscope, provided by the sponsor company, was collected. For recordings as well as for time-performance testing, Nexus 6P (CPU: Qualcomm Snapdragon 810 MSM8994 2.0 GHz octa-core, RAM: 3GB, Camera: 12.3 MP) smartphone was used. Using this ``microscope-smartphone'' setup, video records along Z-axis of 150 specimens from standard AmScope slides were collected. Every specimen Z-stack was recorded in four different speed configurations. For testing fast search of sample focus zone, additional 30 recordings with different speeds and image resolution were collected. For approbation, our mobile microscope library was used as part of Android application with camera capturing using CameraX framework.

\subsection{Fast search of sample focused area}

To evaluate proposed algorithm for fast search of sample focused area, we consider the next metrics. Accuracy (ACC)~--- used to verify that the algorithm is capable of correct peak detection. For each of the available datasets the algorithm was run and the correctness of the peak position and width was determined visually. Precision (PPV) and recall (TPR). One of the stated objectives is not to prune any frames required by the focus stacking algorithm, therefore we have conducted the following, more formal experiment: check that both scenarios, running focus stacking algorithm with in-focus selection on an original stack and on a pruned one, result in the same set of frames. The result of the first scenario is considered ground-truth.

\begin{table}
\centering
\caption{Evaluation of fast search of sample focused area }
    \begin{tabular}{|c|c|c|c|c|c|c|c|}
    \hline
        \textbf{Metric} & ACC & PPV & TPR & T$_{old}$ & T$_{new}$ & SpeedUp \\ \hline
        \textbf{Result} & $0.98$ & $1.00$ & $0.98$ &  $246$\,sec. & $32.8$\,sec. & $7.5$ \\ \hline
    \end{tabular}
\label{peak_search_exprs}
\end{table}

Results of evaluation for collected data is presented in the Table~\ref{peak_search_exprs}. Considered metrics values demonstrate stability of the proposed algorithm in terms of both correct peak detection and correct detection of minimum segment for full focus coverage. Also it drastically speeds up the current process of specimen scanning -- instead of slow scanning of Z-axis that takes almost 5 minutes ($T_{old}$), new strategy of fast scanning for search of focused zone and slow scanning of focused zone takes half of a minute ($T_{new}$).

\subsection{Full focus coverage}
Because the full focus coverage set is not unique due to near-duplicated frames, it is not reasonable to validate algorithm performance using already pre-filtered indices of Z-stack. To evaluate filtering algorithm performance, the next criteria were considered: (i) the total number of frames in the selected stack, (ii) the number of repeated frames among them, (iii) the number of frames that do not contain information about the sample, (iv) the presence of every sample's detail in at least one image in the final stack. The evaluation of the proposed algorithm on the dataset shows that the result stack's size is always near 2-5 frames and doesn't contain duplicates or blurred images. To check the performance of the full focus coverage finding algorithm on the smartphone, three versions of input videos, differing in resolution, were prepared. The result of this performance test is presented in Table~\ref{time_cmp_1}. It can be seen that both methods are quite fast on the low resolution to provide real-time processing for showing a preview of the result just after the end of the recording. Another performance test was done for blur and duplicates filtering of the full focus coverage finding algorithm's output. Its' result is presented in the Table~\ref{time_cmp_duplicates_defocused}. Such filtering takes only a few milliseconds but enriches the quality of full focus coverage set a lot.

\begin{table}
        \centering
        \caption{Performance of full focus coverage finding algorithm, ms, CI=0.95}
        \label{time_cmp_1}
        \begin{tabular}[c]{
    |c| S[table-format=4.4,output-decimal-marker=\times]
    *2{|S[separate-uncertainty=true, table-align-uncertainty=true,
          table-figures-integer=3, table-figures-decimal=2, table-figures-uncertainty=1, 
          table-number-alignment=center]}
    |}
        \hline
          & \multicolumn{1}{c}{Resolution} & \multicolumn{2}{|c|}{Frames selection method} \\ \cline{3-4}
                               & &          \multicolumn{1}{c|}{Parts} & \multicolumn{1}{c|}{Best3} \\ \hline
        1 & 464.848 & 56.25\pm0.5 & 36.73\pm0.1 \\ \hline
        2 & 768.1024 & 156.41\pm1.5 & 126.97\pm1.1 \\ \hline
        3 & 1080.1920 & 615.90\pm7.6 & 472.87\pm3.4 \\ \hline
        \end{tabular}
\end{table}

\begin{table}
    \centering
    \caption{Performance of blur and duplicates filtering on the output of the full focus coverage finding algorithm, ms, CI=0.95}
    \label{time_cmp_duplicates_defocused}
    \begin{tabular}{|c
    *4{|S[separate-uncertainty=true, table-align-uncertainty=true,
          table-figures-integer=2, table-figures-decimal=1, table-figures-uncertainty=1, 
          table-number-alignment=center]}
    |}
    \hline
    & \multicolumn{4}{c|}{Stack size, frames} \\ \cline{2-5}
    & \multicolumn{1}{c}{2} & \multicolumn{1}{|c}{3} & \multicolumn{1}{|c}{4} & \multicolumn{1}{|c|}{5} \\ \hline
    1 & 1.2\pm0.1 & 2.0\pm0.1 & 2.9\pm0.1 & 3.9\pm0.1 \\ \hline
    2 & 2.7\pm0.1 & 3.5\pm0.1 & 5.2\pm0.1 & 6.6\pm0.1 \\ \hline
    3 & 7.0\pm0.1 & 11.4\pm0.1 & 17.0\pm0.1 & 22.3\pm0.2 \\ \hline
    \end{tabular}
\end{table}

\subsection{Focus-stacking}
\begin{figure}
\centering
\includegraphics[width=\linewidth,clip,trim=2cm 2cm 2cm 3cm]{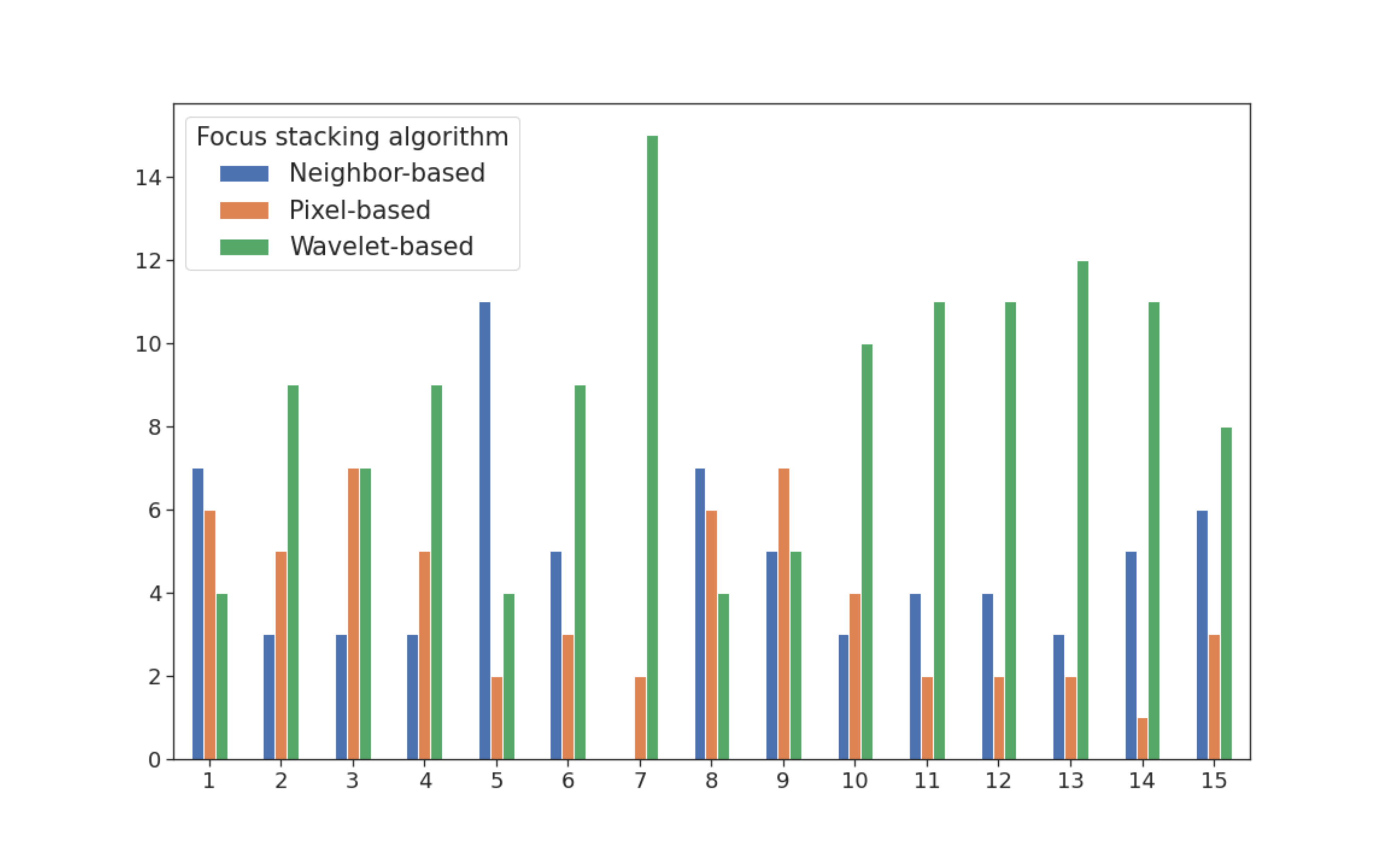}
\caption{Survey results comparing focus stacking results for different algorithms}
\label{survey_desktop}
\end{figure}

Focus-stacking is the final step of the Z-stack processing pipeline, which could be useful in verifying specimen digitization results. Because of there is no ground truth all-in-focus image in the focus-stacking task, we did an anonymous survey with 18 respondents, experts in microscopy imaging, to evaluate the quality of implemented algorithms. A survey was asking to choose one of three images (the results of processing one stack by different algorithms) for 15 sets of images. Images in set 1--5 were taken from a stack of size 2, 1--10 --- stack of size 3, 11--15 --- stack of size more than 3. Figure~\ref{survey_desktop} shows the results of comparison of focus-stacking algorithms on smartphones. According to the survey, the leadership of the wavelet-based focus-stacking algorithm is unambiguous and more noticeable on a longer stack that proves the algorithm's stability. 

For performance measurements, three resolutions of every specimen were selected and specimens with different stack sizes were considered. The results of performance measurements are presented in Table~\ref{time_cmp_fs}. From this point of view, results are opposite~--- wavelet-based algorithm takes more time in comparison to pixel-based and neighbor-based approach, that means they could be used in combination with each other~--- more time-efficient algorithm could demonstrate preliminary result whereas wavelet-based transform could be used for final evaluation of specimen digitization.

\begin{table}
    \centering
    \caption{Performance of pixel-based, neighbor-based, wavelet-based focus stacking algorithms on stacks of different sizes, CI=0.95 }
    \begin{tabular}{| S[table-format=4.4,output-decimal-marker=\times]
    *4{|S[separate-uncertainty=true, table-align-uncertainty=true,
          table-figures-integer=4, table-figures-decimal=1, table-figures-uncertainty=3, 
          table-number-alignment=center]}
    |}
    \hline
    \multicolumn{1}{|c}{Resolution} & \multicolumn{4}{|c|}{Stack size, frames} \\ \cline{2-5}
     & \multicolumn{1}{c}{2} & \multicolumn{1}{|c}{3} & \multicolumn{1}{|c}{4} & \multicolumn{1}{|c|}{5} \\ \hline
        \multicolumn{5}{|c|}{Pixel-based approach, ms} \\ \hline
        464.848 & 30.0\pm0.2 & 42.8\pm0.4 & 49.9\pm1.0 & 63.1\pm1.4  \\ \hline
        768.1024 & 56.2\pm0.3 & 78.8\pm0.5 & 112.7\pm1.1 & 128.2\pm2.9 \\ \hline
        1080.1920 & 150.6\pm2.2 & 222.7\pm3.6 & 286.8\pm3.7 & 415.5\pm8.6  \\ \hline
        \multicolumn{5}{|c|}{Neighbor-based approach, ms} \\ \hline
        464.848 & 355.9\pm3.9 & 340.1\pm9.3 & 363.4\pm4.6 & 457.9\pm11.8 \\ \hline
        768.1024 & 575.4\pm5.5 & 773.0\pm4.7 & 853.2\pm8.6 & 870.7\pm27.6 \\ \hline
        1080.1920 & 1877.9\pm20.1 & 2544.1\pm27.1 & 2952.8\pm32.2 & 3487.0\pm44.8 \\ \hline
        \multicolumn{5}{|c|}{Wavelet-based approach, sec} \\ \hline
        464.848 & 2.5\pm0.1 & 3.2\pm0.1 & 3.8\pm0.1 & 4.7\pm0.1 \\ \hline
        768.1024 & 4.4\pm0.1 & 6.1\pm0.1 & 9.0\pm0.1 & 9.6\pm0.2 \\ \hline
        1080.1920 & 24.2\pm0.2 & 24.3\pm0.3 & 30.5\pm0.1 & 37.4\pm0.2 \\ \hline
    \end{tabular}
    \label{time_cmp_fs}
\end{table}

\section{Conclusion}

In this paper we have presented a concept of a ``smart'' mobile microscope~--- the first platform to combine a mobile device, a microscope and a set of computer vision methods to provide means of specimen digitization, low-dependent of operator's qualification. We have reviewed existing methods in such sub-tasks of Z-stack digitization as: search of sample focused zone, extraction of full focus coverage subset, and focus-stacking for visual verification of correct Z-stack processing. We have demonstrated that fast search of sample focused zone can speed up the process of information capturing more than 8 times without reducing the final image set quality. Additionally, we have shown that Z-stack image filtering for full focused coverage extraction can be done with near real-time performance on a smartphone. Finally, we have discussed modifications of focus-stacking algorithm, estimated their visual quality on the data obtained from the mobile microscope and measured time-performance on target devices. From the software engineering point of view, we have presented design and architecture of a mobile microscope library and demonstrated its cross-platform compatibility and time-efficiency on a variety of smartphones. The experimental results have shown that the proposed mobile library and algorithms are capable of solving the stated tasks while maintaining adequate response time. Thus, a justification for further development of the tool is given.

\section{Aknowledgments}

Authors would like to express their gratitude to MEL Science company for providing equipment for the experiments: mobile microscope and smartphones.

\bibliographystyle{bibtex/spmpsci}
\bibliography{bibtex/egbib.bib}

\end{document}